\documentstyle[agupp,epsf]{article}
\lefthead{LANDGRAF}
\righthead{MODELING INTERSTELLAR DUST DISTRIBUTION}
\authoraddr{M. Landgraf, 
NASA/Johnson Space Center, 
Mailcode: SN2,
2101 NASA Road 1,
Houston, TX 77058, U.S.A.
(e-mail: Markus.K.Landgraf1@jsc.nasa.gov)}
\slugcomment{To appear in the {\it Journal of Geophysical
Research} (JGR special section: ``Interstellar Dust and the
Heliosphere''), invited contribution}
\begin{document}
\title{Modeling the Motion and Distribution of Interstellar Dust inside the
Heliosphere}
\author{M. Landgraf}
\affil{NASA/Johnson Space Center, Houston, Texas, U.S.A.}
\begin{abstract}
The interaction of dust grains originating from the local interstellar
cloud with the environment inside the heliosphere is investigated. As
a consequence of this interaction the spatial distribution of
interstellar dust grains changes with time. Since dust grains are
charged in the interplanetary plasma and radiation environment, the
interaction of small grains with the heliosphere is dominated by their
coupling to the solar wind magnetic field. The change of the field
polarity with the solar cycle imposes a temporal variation of the
spatial distribution and the flux of small (radius smaller than
$0.4\;{\rm \mu m}$) interstellar dust grains in the Solar System,
whereas the flux of large grains is constant because of their
negligible coupling to the solar wind magnetic field. The flux
variation observed by in-situ measurements of the Galileo and Ulysses
spacecraft are reproduced by simulating the interaction of
interstellar grains with charge-to-mass ratios between $0.5\;{\rm
C}\;{\rm kg}^{-1}$ and $1.4\;{\rm C}\;{\rm kg}^{-1}$ with the
interplanetary environment.
\end{abstract}
\begin{article}
\section{Introduction}
Levy and Jokipii [1976] showed that the dominant force on small dust
grains in the Solar System is the Lorentz-force caused by the solar
wind magnetic field sweeping by the dust grains, which are charged to
a positive surface potential, mainly by electron emission due to solar
UV photo-effect. It was concluded that due to the coupling to the
radially outward flowing solar wind, small interstellar grains do not
exist inside the heliosphere which defines the domain of the solar
wind. For big grains, Lorentz-force is less important, because they
have smaller charge-to-mass ratios, but from the determination of the
size range and size distribution of interstellar grains in our galaxy
by fitting the wavelength dependence of extinction of starlight on its
way through the interstellar medium [e.g. {\em Mathis et al.}, 1977],
it was believed that typical interstellar grains are not larger than
$0.5\;{\rm \mu m}$ in diameter. So the conclusion was drawn that there
are no big interstellar dust grains and the small ones are swept out
of the Solar System by the solar wind magnetic field. Gustafson and
Misconi [1979] on the other hand, argued that the electromagnetic
interaction enhances the spatial density of small (radii between
$0.0005\;{\rm\mu m}$ and $0.12\;{\rm \mu m}$) interstellar grains
upstream of the Sun. The detection of interstellar grains in the Solar
System by the dust detector on-board Ulysses [\markcite{{\it Gr\"un et
al.,} 1993}] and its confirmation by data from the Galileo dust
detector [\markcite{{\it Baguhl et al.,} 1995}] proved the existence of
grains larger than $0.5\;{\rm \mu m}$ and that smaller grains can
penetrate into the Solar System, although they are depleted in number
[\markcite{{\it Gr\"un et al.,} 1994}]. It was concluded
[\markcite{{\it Gr\"un et al.,} 1994}; \markcite{{\it Gustafson and
Lederer,} 1996}; \markcite{{\it Grogan et al.,} 1996}]
that the Lorentz-force on interstellar grains not only repels but also
focuses them, depending on the configuration of the solar wind
magnetic field, and thus on the phase of the solar cycle. The
objective of this work is to quantitatively determine the spatial
distribution of interstellar dust grains in the Solar System as a
function of the solar cycle. To meet this objective, the motion of
interstellar grains in the Solar System is simulated by numerically
integrating the equation of motion for a big ($N = 10^6$ to $10^7$)
set of interstellar grains. In section \ref{sec_modeldescription} the
assumptions and approximations used in the model are
described. Section \ref{sec_results} presents the resulting
three-dimensional and time-dependent spatial distributions of grains
in the Solar System. The results are discussed and compared with the
in-situ measurements of the Ulysses and Galileo spacecraft in section
\ref{sec_discussion}.

\section{\label{sec_modeldescription}Model Description}
As initial conditions, a homogeneous, mono-di\-rec\-tion\-al stream of
grains is assumed arriving from an upstream direction of $\lambda_{\rm
ecl} = 259^\circ$ (heliocentric ecliptic longitude) and $\beta_{\rm
ecl} = +8^\circ$ (heliocentric ecliptic latitude). This direction is
equal to the upstream direction of interstellar dust grains in the
Solar System as it was determined by analyzing the directional
information of the Ulysses in-situ measurements [\markcite{{\it Baguhl
et al.,} 1995}; \markcite{{Frisch et al.,} 1999}]. This direction is
also compatible with the upstream direction of interstellar neutral
helium atoms, determined by the measurements of the Ulysses/GAS
experiment [\markcite{{\it Witte et al.,} 1996}]. It was shown
[\markcite{{\it Gr\"un et al.,} 1994}], that the impact velocity of
interstellar grains measured by Ulysses after Jupiter fly-by was
compatible with the velocity of the interstellar helium atoms. Because
the impact velocity of an individual grain is determined with an
uncertainty of factor $2$ by the in-situ measurement [\markcite{{\it
Gr\"un et al.,} 1992}],
the helium velocity of $v_\infty=25.3\;{\rm km}\;{\rm s}^{-1}$ was
adopted for the model. In the simulation, a ``wall'' of grains is
placed at a heliocentric distance of $100\;{\rm AU}$ every time
interval $\Delta t$. The wall covers an area of $30\;{\rm AU}\times
30\;{\rm AU}$ for big and $120\;{\rm AU}\times 200\;{\rm AU}$ for
small grains, and is $v_\infty\Delta t$ thick. Initially the grains
are distributed uniformly inside the wall.

The position $\vec{x}$ and velocity $\dot{\vec{x}}$ of each grain as a
function of time is determined by integrating the equation of motion
given by:
\begin{eqnarray}
\ddot{\vec{x}} + \left(1 - \beta \right) \frac{G M_\odot} {\left|
\vec{x} \right|^3}\vec{x} - \frac{q}{m} \left(\left(\dot{\vec{x}} -
\vec{v}_{\rm sw}\right) \times
\vec{B}_{\rm P}\right) & = & 0\label{dgl},
\end{eqnarray}
where $\beta$ is the strength of radiation pressure expressed as the
ratio of the magnitude of radiation pressure force to the magnitude of
gravity, $G$ is the gravitational coupling constant, $M_\odot$ the
mass of the Sun, $q$ and $m$ the charge and mass of the grain,
$\vec{v}_{\rm sw}$ the solar wind velocity, and $\vec{B}_{\rm P}$ the
solar wind magnetic field. In the case $q/m \rightarrow 0$, i.e. if
the Lorentz-force on the grains is negligible, the spatial
distribution of grains can be determined analytically
[\markcite{{\it Landgraf and M\"uller,} 1998}] and the solution is
rotationally symmetric about
the axis parallel to the upstream direction.
If the Lorentz-force can not be neglected, the solution to equation
(\ref{dgl}) has no continuous symmetry and the dust grain trajectories
are essentially three-dimensional.  Equation (\ref{dgl}) contains
three basic parameters: the strength of radiation pressure $\beta$, the
charge-to-mass ratio $q/m$, and the strength of the radial $B_{r,0}$
and azimuthal $B_{\phi,0}$ components of the solar wind
magnetic field.
To reduce the number of free parameters in the model, spherical grains
with a constant density of $\rho = 2.5\;{\rm g}\;{\rm cm}^{-3}$ are
assumed. In this case, $\beta$ and $q/m$ can be expressed as a
function of grain radius $a$. In the model the calculation of $\beta$
as a function of grain radius by Gustafson [1994] is used (see also
table \ref{tab_cycle} below).  In this
Mie-theory calculation, the optical constants of ``astronomical
silicates'' introduced by Draine and Lee [1984] are assumed. The
parameter $\beta$
reaches its maximum for $a = 0.18\;{\rm \mu m}$ and is greater than
unity for $0.09\;{\rm \mu m} < a < 0.4\;{\rm \mu m}$. In this regime,
interstellar grains are effectively repelled by the Sun. For
$a\rightarrow 0$, $\beta$ reaches a constant value of $\approx 0.1$,
and for $a\rightarrow \infty$, $\beta$ depends on $a$ like $a^{-1}$,
which is valid in the limit of geometric scattering.

For the charge on a dust grain in interplanetary space it is assumed
that the charging process due to photo-emission of electrons induced
by solar UV photons partially balanced by the interplanetary plasma
environment produces a positive surface potential of $U = +5\;{\rm V}$
on the grain surface [\markcite{{\it Gr\"un et al.,} 1994}]. The charge of the grain is then
given by
\begin{eqnarray}
q & = & 4\pi\epsilon_0 U a\nonumber,
\end{eqnarray}
where $\epsilon_0$ is the permeability of the vacuum. Thus, $q/m
\propto a^{-2}$, which means that electromagnetic interaction with
the solar wind magnetic field is only important for small grains.

In the model, Parker's [1958] model is used, which has been
found to be a good large scale approximation of the observed magnetic
field in interplanetary space [\markcite{{\it Mariani and Neubauer,}
1990}; \markcite{{\it Burlaga and Ness,} 1993}; \markcite{{\it Smith
et al.,} 1995}]. The
radial, azimuthal, and normal components ($B_r$, $B_\phi$, $B_\theta$)
of the Parker-field can be expressed as a function of location in
heliographic coordinates ($r_{\rm hg}$, $\lambda_{\rm hg}$,
$\beta_{\rm hg}$):
\begin{eqnarray}
B_r & = & \pm B_{r,0}\left(\frac{r_0}{r_{\rm hg}}\right)^2\nonumber\\
B_\phi & = & \pm B_{\phi,0} \frac{r_0}{r_{\rm hg}} \cos \beta_{\rm hg}
\label{eqn_parkerfield}\\
B_\theta & = & 0\nonumber,
\end{eqnarray}
where $B_{r,0} = B_{\phi,0} = 3\;{\rm nT}$ are the radial and
azimuthal field strength at $r_0 = 1\;{\rm AU}$, respectively
[markcite{{\it Gustafson,} 1994}]. The Lorentz-force $\vec{F}_{\rm L}$
on a dust
grain in interplanetary space caused by the solar wind magnetic field
is dominated by the effect of the field sweeping by the grain with the
solar wind speed, because the grain velocity is much smaller than the
solar wind velocity. Therefore, the Lorentz-force is governed by the
term $(q/m)(B_{\phi}v_{\rm sw})$ (see equation (\ref{dgl})). Although
it would seem that the Lorentz-force depends on the solar wind speed, this
is not the case, because at higher solar wind speeds, the Parker
spiral becomes steeper, and consequently the azimuthal component
$B_\phi$ is reduced [\markcite{{\it Gustafson and Misconi,} 1979}],
effectively keeping the
Lorentz-force constant. The observed sector structure of the polarity
of the solar wind magnetic field has been modeled by Alfven [1977]
using a ``ballerina model''. This model describes a warped current
sheet that separates opposite polarities of the magnetic field in
interplanetary space. In the ecliptic, opposite polarities are
observed in separate sectors, the boundaries of which are given by the
intersections of the current sheet with the ecliptic plane. For dust
grains in the considered size-range between $a=0.1\;{\rm \mu m}$ and
$a = 0.7\;{\rm \mu m}$, the instantaneous polarity can be replaced by an
average polarity $\overline{p}$ which is a function of the
heliographic latitude $\beta_{\rm hg}$. This averaging process
is a good approximation for grains for which the Larmor-frequency is
orders of magnitude smaller than the solar rotation frequency
$\omega_\odot$, therefore,
\begin{eqnarray}
a & \gg & \sqrt{ \frac{3 \epsilon_0 U B_0} {\rho \omega_\odot}} =
8\;{\rm nm},\nonumber
\end{eqnarray}
where $B_0 = 3\;{\rm nT}$ is the interplanetary magnetic field strength
at $1\;{\rm AU}$ and $\omega_\odot = 2.7\cdot 10^{-6}$ is the solar
rotation frequency (rotational period of 27 days). The effect of the
sector structure on very small ($6\;{\rm nm}$ to $10\;{\rm nm}$) dust grains
which have been ejected by the Jovian system, has been observed by Ulysses
[\markcite{{\it Zook et al.,} 1996}]. The average polarity
$\overline{p}$ seen by a dust
grain at heliographic latitude $\beta_{\rm hg}$, is given by
\begin{eqnarray}
\overline{p} & = & \frac{2}{\pi}\arcsin\frac{\tan \beta_{\rm hg}}
{\tan \epsilon}\label{eqn_mean_polarity}\\
B_{r,\phi} & \rightarrow & \overline{B}_{r,\phi} = \overline{p} \left|
B_{r,\phi} \right|\nonumber,
\end{eqnarray}
where $\epsilon$ is the tilt angle of the (equivalent) flat current
sheet, that rotates with the $22$-year period of the solar cycle
[\markcite{{\it Gustafson,} 1994}]. From equation (\ref{eqn_mean_polarity}) it is
clear that the {\em average} field strength $\overline{B}_{r,\phi}$ is
small at low heliographic latitudes, whereas the {\em instantaneous}
field strength is high at low heliographic latitudes (see equation
(\ref{eqn_parkerfield})). This is because a particle at low latitudes
experiences a negative polarity nearly as long as a positive
polarity. It was shown [\markcite{{\it Morfill and Gr\"un,} 1979}]
that the stochastic
perturbation of the trajectories of small dust grains by the sectored
solar wind magnetic field produces an outward directed diffusion of
these grains. This is not taken into account by the model, because the
sector structure was removed by using an average polarity.

The average polarity $\overline{p}$ defines the sign of the azimuthal
component of the average magnetic field and thus the sign of the
Lorentz-force, and because the average polarities in the northern and
southern hemispheres are always opposite, the Lorentz-force
accelerates dust grains away from the solar equatorial plane for
$\overline{p} > 0$ and towards this plane for $\overline{p} < 0$. The
sequence of this focusing/defocusing cycle is shown in
{table~1}.

To simulate the motion of interstellar grains in the heliosphere,
equation (\ref{dgl}) was integrated by using a simple
Runge-Kutta integrator of fourth order [\markcite{{\it Numerical
Recipes,} 1992}]. After the
first grains have completely traversed the heliosphere, the spatial
distribution of dust grains has been determined at each time step by
counting grains inside each cell of an orthogonal, equidistant grid.
The resulting spatial distributions of interstellar grains in the
heliosphere are discussed in the following sections.

The model described above uses some non-trivial
approximations. For the determination of the basic parameters $\beta$
and $q/m$, a spherical grain shape has been assumed. On the other
hand, the observation of the polarization of starlight (for summary
see Li and Greenberg [1997]) has
shown that interstellar dust grains (at least the part of the
population that causes the extinction) must have elongated shapes. A
deviation from a spherical shape would increase $q/m$ for the same
surface potential $U$, because the charge carriers on the grains are
separated by larger distances. The $\beta$-value would also increase
for non-spherical particles, because it is proportional to the
cross-section-to-volume ratio which is minimal for spheres. It can
therefore be concluded that a non-spherical particle of the same size
as a spherical one has larger values for the basic parameters $q/m$
and $\beta$, or, a non-spherical particle with the same $q/m$ and
$\beta$ values used in the simulation is larger than indicated in
table \ref{tab_dynpara}. Still, $q/m$ and $\beta$ are quantities
that depend on material properties of the grains and have not been
measured directly. Therefore, they have to be treated as free
parameters.

Another assumption made in the model is the constancy of $q$. Due to
short-term variations in the solar wind, the surface potential and thus
$q$ is expected to fluctuate. These charge fluctuations have the same
effect on the grain's motion as short-term variations of the ambient
magnetic field. As shown above, it is a good approximation to use an
average strength of the Lorentz-force, with the averaging time-scale
being the Larmor-period of the grain. Furthermore it was shown by Zook
et al. [1996], who modeled the motion of very small Jupiter stream
particles through interplanetary space, that the effect of charge
fluctuations on grain dynamics can be neglected for the grain sizes
under consideration.

Finally it can be argued that the solar wind magnetic field is not
well represented by a Parker spiral during the solar maximum. The
physical situation during solar maximum is a highly disordered field
with the polarity distribution showing no clear separation of
polarities between the northern and southern heliographic hemisphere
[\markcite{{\it Hoeksema,} 1985}]. As shown above, the motion of
grains in the size
range of $0.1\;{\rm \mu m}$ depends only on the {\em average} polarity
$\overline{p}$. In the highly disordered configuration of the solar
maximum, regions of positive and negative polarity act on the grains
in a stochastic way, independent of the heliographic latitude of the
grain's position. Therefore $\overline{p}=0$ and thus vanishing
average Lorentz-force for all heliographic latitudes is a good
approximation of the physical situation during the solar maximum,
independent of the actual form of the field lines.

\section{Spatial Distributions of Interstellar Dust in the Solar
System\label{sec_results}} 
\begin{figure}[ht]
\centering
\epsfbox{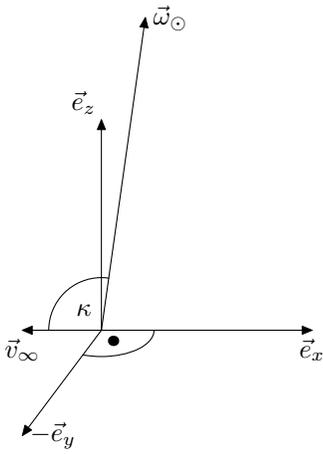}
\caption{\label{fig_coorsys}Coordinate-system in which the spatial
distribution of interstellar dust in the heliosphere are displayed. The
$x$-$y$-plane is close ($<10^\circ$ off) to the cliptic. A tilt angle
$\kappa$ of $90^\circ$ between the downstream-direction and the solar
rotation axis is compatible with the determination of the
stream-direction, but the best fit to the in-situ data is achieved at
$\kappa=98^\circ$ [\markcite{{\it Frisch et al.,} 1999}].}
\end{figure}

The spatial distribution of interstellar
grains inside the heliosphere shows how the concentration of grains of
different sizes is reduced or enhanced by solar gravity, radiation
pressure, and the solar wind magnetic field. To display the
distributions that result from the model described in section
\ref{sec_modeldescription}, a coordinate-system is chosen such that
the $x$-axis is anti-parallel to the initial velocity vector. The
$z$-axis is then chosen to be perpendicular to the $x$-axis and to lie
in the plane of the initial velocity vector $\vec{v}_\infty$ and the
solar rotation axis vector $\vec{\omega}_\odot$ (see
{figure~\ref{fig_coorsys}}). The $y$-axis is set to be
perpendicular to the $x$- and $z$-axes such that a right-handed
coordinate-system is formed.

The angle $\kappa$ between the downstream-direction and the solar
rotation axis (as shown in figure \ref{fig_coorsys}), has been
determined by fitting the directional information of the Ulysses
in-situ measurements. The best fit was achieved at $\kappa=98^\circ$
[\markcite{{\it Frisch et al.,} 1999}].

The result of the simulation is shown for four different grain radii:
$0.7\;{\rm \mu m}$, $0.5\;{\rm \mu m}$, $0.3\;{\rm \mu m}$, $0.2\;{\rm
\mu m}$, and $0.1\;{\rm \mu m}$. For the dynamical parameters $\beta$
and $q/m$ of these grains see
{table~2}.

The spatial distribution of the $0.7\;{\rm \mu m}$ and $0.5\;{\rm \mu
m}$ grains is shown in {figure~\ref{fig_dd0705}}. Because of
the small charge-to-mass ratio of these grains, the coupling to the
time-variable solar wind magnetic field is weak, and therefore their
spatial distribution is nearly constant and rotationally symmetric
about the $x$-axis (downstream direction). The density enhancement
downstream of the Sun is produced by gravitational bending of the
grain's trajectories towards the Sun. Due to their smaller
$\beta$-value, the downstream concentration increase of the
$0.5\;{\rm \mu m}$-grains is less pronounced. The comparison with the
result of the analytical calculation [\markcite{{\it Landgraf and
M\"uller,} 1998}], which assumes
$q/m = 0$, shows that electromagnetic interactions can indeed be
neglected for these particles. The spatial distribution of grains with
radii of $0.3\;{\rm \mu m}$ is already slightly affected by their
interaction with the solar wind magnetic field.

For even smaller grains the interaction with the solar wind magnetic
field becomes important, as shown in
{figure~\ref{fig_dd02_xy}}, in which the time-sequence of the
density distribution of interstellar $0.2\;{\rm \mu m}$-grains in the
$x$-$y$-plane (approximately in the ecliptic plane) is displayed. As
shown in table~1, the solar maximum 1991 was the end of a focusing
cycle. The density enhancement in the $x$-$y$-plane is the result of
the focusing of grains to this plane. At the maximum of solar activity
the average magnetic field is weak, because the current sheet extends
to high heliographic latitudes. In this configuration sectors of
positive and negative polarity cancel each other efficiently as they
pass by the dust grains, creating a weak average
Lorentz-force. Therefore, the region of enhanced grain density moves
downstream without much perturbation until in 1996 the defocusing
effect becomes important. The void region downstream of the Sun is
expanded by the defocusing, and the overall density in the
$x$-$y$-plane is reduced until the end of the defocusing cycle in
2002. After that the new focusing cycle starts to increase the density
in the $x$-$y$-plane until the distribution in 2012 is similar to the
one in 1991 when the defocusing cycle started.  As a consequence of
the large $\beta$-value of $\beta=1.4$, the concentration of
$0.2\;{\rm \mu m}$-grains downstream of the Sun is very low.  In the
$x$-$z$-cut (plane approximately perpendicular to the ecliptic that
also contains the stream vector) of the density distribution shown in
{figure~\ref{fig_dd02_xz}} one can see that the density
distribution is no longer symmetric about the $x$-axis for $0.2\;{\rm
\mu m}$-grains. The first panel in figure \ref{fig_dd02_xz} also
shows that because the upstream direction lies close to but not in the
plane of the solar equator, the distribution is asymmetric about the
$x$-$y$-plane. At the end of the focusing cycle
in 1991 the grains are concentrated along the plane of the solar
equator and the density at higher latitudes is reduced. As the
defocusing cycle proceeds, the density at higher latitudes increases
as more grains are deflected into this region. In 2002, at the end of
the defocusing cycle the regions of enhanced density move downstream
and the new focusing cycle starts again to concentrate grains around
the plane of the solar equator until in 2012 the distribution of 1991
is restored.

The spatial distribution of grains with radii of $0.1\;{\rm \mu m}$ is
dominated by the interaction of these grains with the solar wind
magnetic field due to their high charge-to-mass ratio of $q/m =
5.31\;{\rm C}\;{\rm kg}^{-1}$. At the end of the focusing solar cycle,
the grains are concentrated in a sheet about the plane of the solar
equator as can be seen in the first panel in
{figure~\ref{fig_dd01_xz}}. Because the solar wind magnetic
field moves radially outward with the solar wind, the Lorentz-force on
the grains is not perfectly perpendicular to the solar equator, but
also contains a small outward directed component. This force
decelerates small grains and causes a region of enhanced density
upstream of the Sun, which is visible as the arc-shaped feature in the
first panel of {figure~\ref{fig_dd01_xy}}. As the average
field strength is low during the solar maximum, this feature can
propagate towards the Sun and the concentration around the plane of
the solar equator is split into a northern and southern part by the
new defocusing field (1993 to 2001 in figure \ref{fig_dd01_xz}). As
can be seen in the panel for 1996 in figure \ref{fig_dd01_xy}, the
region of high grain density is diverted before it reaches the solar
vicinity. Therefore, the flux of $0.1\;{\rm \mu m}$ grains measured in
the solar vicinity (roughly inside Jupiter's orbit) is expected to be
reduced during the full solar cycle due to the interaction with the
solar wind magnetic field. This deficiency of small grains has been
observed in the in-situ data [\markcite{{\it Gr\"un et al.,} 1994};
\markcite{{\it Landgraf et al.,} 1999}]. After a new
focusing cycle starts in 2002, grains are again concentrated around
the solar equator as can be seen in the panels from 2006 to 2012 in
figure \ref{fig_dd01_xz} until the configuration of 1991 is restored.

The simulation shows that the flux of small interstellar grains in the
solar vicinity depends on the phase of the solar cycle. Since in the
interstellar medium, small grains ($a < 0.2\;{\rm \mu m}$) dominate
the dust population by number as indicated by extinction measurements
[\markcite{{\it Mathis et al.,} 1977}; \markcite{{\it Kim et al.,}
1994}], the total flux of interstellar grains in the
Solar System should be dominated by smallest grains that are not
filtered by the solar wind magnetic field. Therefore, the total flux
of interstellar dust depends on the phase of the solar cycle. In the
next section the flux of interstellar grains measured in-situ with the
Galileo and Ulysses dust detectors is compared with the temporal
variation predicted by the model.

\section{Comparison with In-Situ Measurements\label{sec_discussion}}
The magnitude $f_{\rm isd}(t_1,t_2)$ of flux of interstellar dust measured by
the dust detectors on-board Ulysses and Galileo in the time-interval
$[t_1,t_2]$ is determined by $f_{\rm isd}=N_{\rm isd}(t_1,t_2)/I(t_1,t_2)$,
where $N_{\rm isd}(t_1,t_2)$ is the number of interstellar grains
detected in the given time-interval and $I(t_1,t_2)$ is the
average sensitive detector-area, accumulated between $t_1$ and
$t_2$. $I(t_1,t_2)$ is given by:
\begin{eqnarray}
I(t_1,t_2) & = & \int_{t_1}^{t_2} dt \int_0^{2\pi}\frac{d\phi_{\rm
rot}} {2\pi} A(t,\phi_{\rm rot}) \frac{v_{\rm rel}(t)} {v_{\infty}}
\nonumber,
\end{eqnarray}
where $\phi_{\rm rot}$ is the spacecraft rotation angle (Ulysses and
Galileo are spin-stabilized spacecraft, and therefore the sensitive
area for a 
dust stream from one direction has to be averaged over the rotation
angle), $A$ is the sensitive detector-area of the dust detector which
is a function of time and rotation angle, $v_{\infty}$ is the dust
velocity at large distances from the Sun, and $v_{\rm rel}$ is the
relative velocity of the spacecraft and the dust as a function of
time. By introducing the factor $v_{\rm rel}/v_\infty$, the effect of
the flux measured in the spacecraft frame being enhanced if
the spacecraft moves upwind and reduced during downwind
motion is taken into account. Therefore, $f_{\rm isd}$ is the flux in
the inertial frame at the location of the spacecraft at time $t$.

\begin{figure}[ht]
\begin{tabular}{cc}
\epsfxsize=.8\hsize
\epsfbox{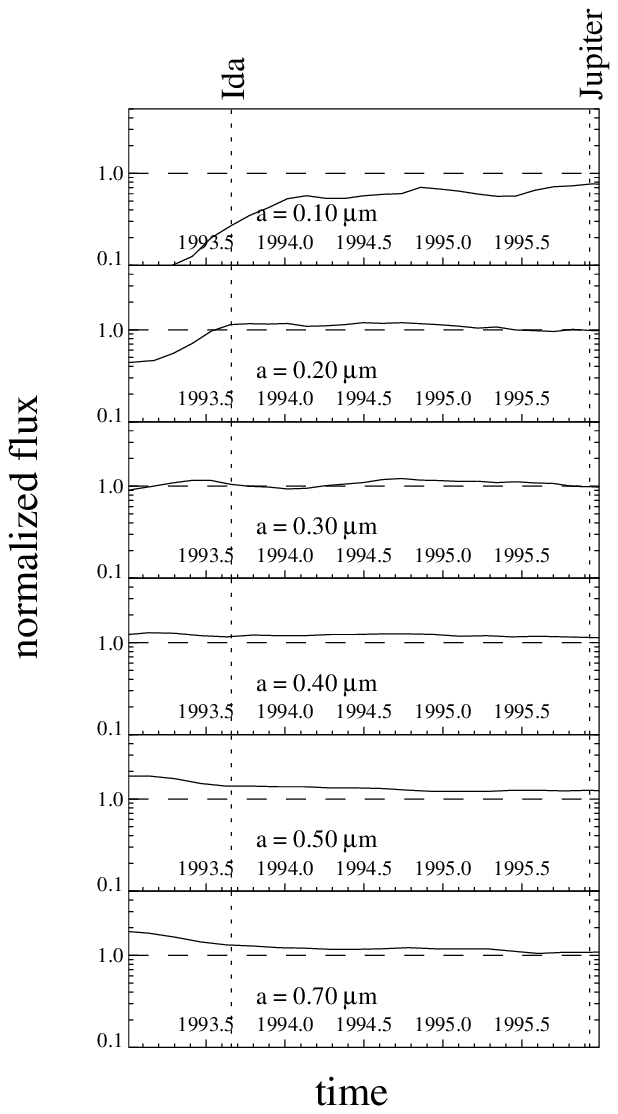} &
\epsfxsize=.8\hsize
\epsfbox{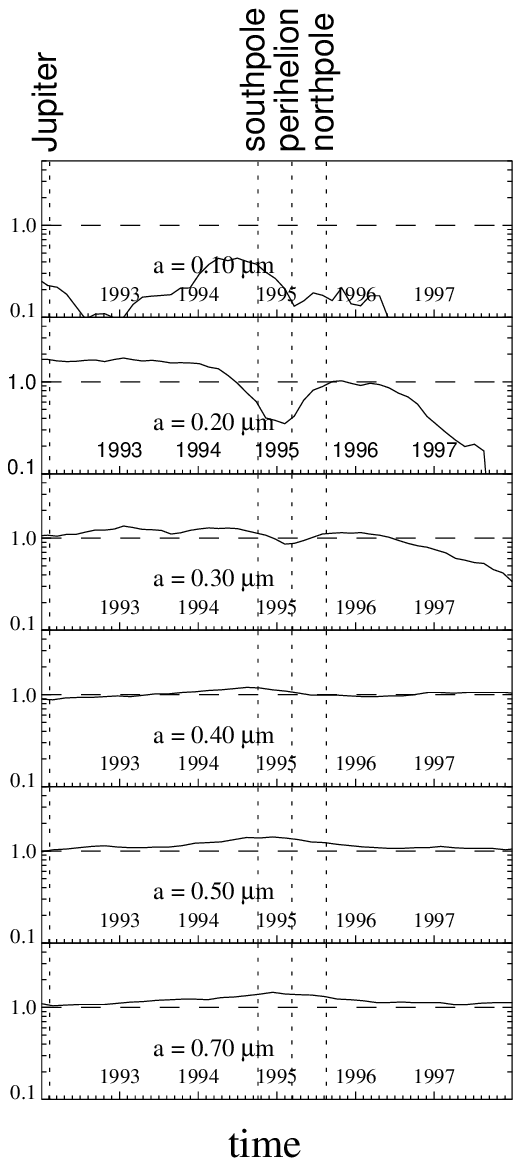}
\end{tabular}
\caption{\label{fig_modelfluxes} The flux of interstellar dust in the
inertial frame at the location of Galileo (left panel) and Ulysses
(right panel) as a function of time as predicted by the
simulation. The magnitude of the flux is normalized to its value at
large distances from the Sun. The radius $a$ of the simulated grains
increases from the top to the bottom. Major mission events are
indicated by dotted vertical lines. In both cases, the flux of small
grains varies stronger than the flux of big grains due to their
coupling to the solar wind magnetic field. The flux of $0.1\;{\rm \mu
m}$-grains strongly reduced at Ulysses locations.}
\end{figure}

In the following $f_{\rm isd}(t)$ predicted by the model is
compared with the flux measured with the Ulysses and Galileo
spacecraft. For the comparison it is assumed, that the population of
interstellar dust grains is dominated by grains out of a narrow
size-interval which can be represented by one of the simulated
grain-sizes (and thus dynamical parameters $\beta$ and $q/m$). In
{figure~\ref{fig_modelfluxes}} the model
prediction for the flux of 
interstellar grains at the location of the Ulysses and Galileo
spacecraft is shown as a function of time.

For the Galileo measurements the simulation predicts a constant flux
of interstellar grains on the detector for all but the smallest grain
sizes. This is because of the short period of time covered by the
Galileo data. Interstellar grains could be identified clearly in the
Galileo data only after it left the inner Solar System in mid-1993 and
before it reached the Jovian system at the end of 1995. This time
interval is short compared to the duration of the solar cycle,
therefore the change of the solar cycle did not affect the flux of
$0.2\;{\rm \mu m}$- and $0.3\;{\rm \mu m}$-grains. The predicted
increase in the flux of $0.1\;{\rm \mu m}$-grains is due to to
Galileo's upstream directed motion towards the region of enhanced
spatial density as can be seen in the first three panels of figure
\ref{fig_dd01_xy}.

The prediction for the long-term measurements of Ulysses from February
1992 (Jupiter flyby) to end of 1997 (latest data) shows that for
$0.1\;{\rm \mu m}$-, $0.2\;{\rm \mu m}$, and $0.3\;{\rm \mu m}$-grains
the effects of the solar cycle should modulate the measured flux. In
the Ulysses data, the flux of $0.1\;{\rm \mu m}$-grains is predicted to
be strongly reduced. After 1996 the simulation predicts a decrease in
the flux of the $0.2\;{\rm \mu m}$- and $0.3\;{\rm \mu
m}$-grains. This is due to the defocusing effect of the solar wind
magnetic field on the trajectories of these grains. The flux of larger
grains is nearly constant and close to its value at large distances,
because they couple weakly to the solar wind magnetic field.

\begin{figure}[h]
\epsfxsize=2\hsize
\epsfbox{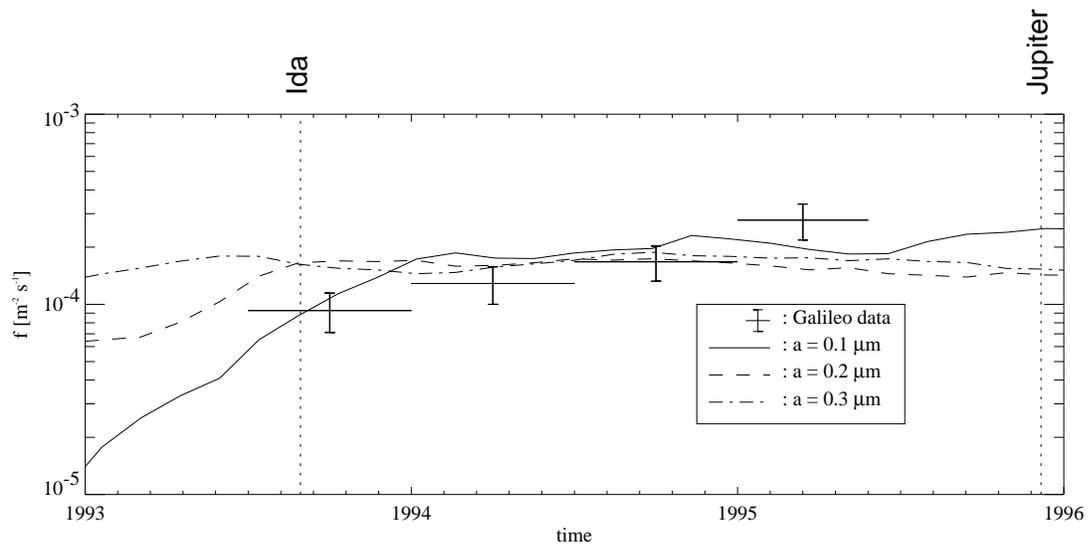}
\caption{\label{fig_gal_flux_comp} Comparison of the temporal
variation of the flux in the inertial frame at Galileo's location
predicted by the simulation (solid line for $0.1\;{\rm \mu m}$-grans,
dashed line for $0.2\;{\rm \mu m}$-grans, and dash-dotted linefor
$0.3\;{\rm \mu m}$-grans) with Galileo measurements (crosses). Due to
the short time interval covered by the measurements, no significant
changes of the flux due to the solar cycle are expected.}
\end{figure}

\begin{figure}[h]
\epsfxsize=2\hsize
\epsfbox{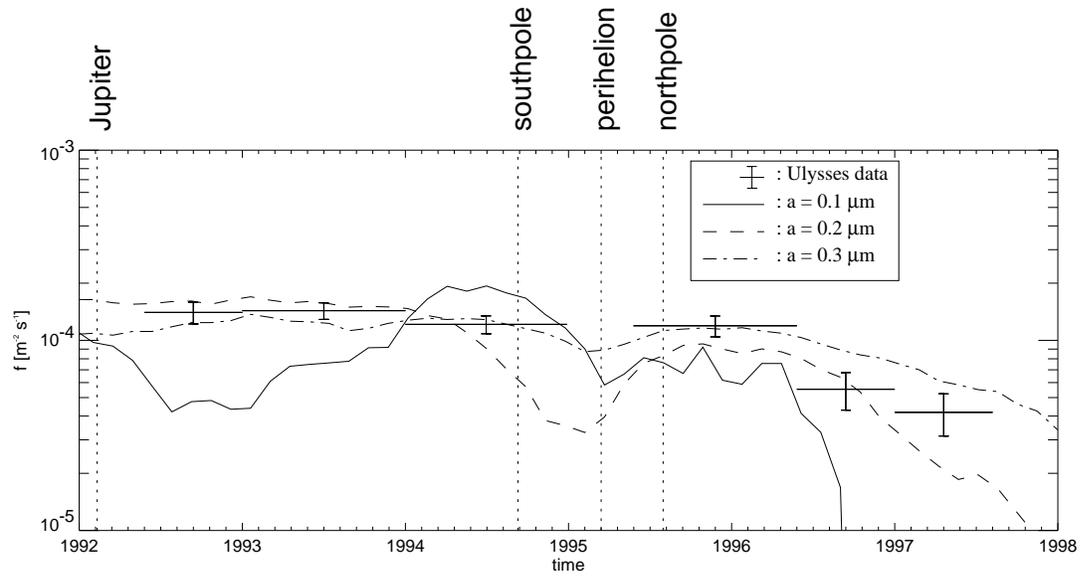}
\caption{\label{fig_uls_flux_comp}Comparison of the temporal
variation of the flux in the inertial frame at Ulysses' location
predicted by the simulation (solid line for $0.1\;{\rm \mu m}$-grans,
dashed line for $0.2\;{\rm \mu m}$-grans, and dash-dotted linefor
$0.3\;{\rm \mu m}$-grans) with Ulysses measurements (crosses). Because
impacts of interstellar grains could not be distinguished clearly from
impacts of interplanetary grains around perihelion passage, there is a
gap in the measured flux between January and May 1995. The measured
decrease in flux after 1996 can be due to the onset of the defocusing
cycle in 1991 as indicated by the results of the simulation.}
\end{figure}

{Figures~\ref{fig_gal_flux_comp}}
{and~\ref{fig_uls_flux_comp}} show the comparison of the
temporal variation of the flux predicted by the simulation with the
Galileo and Ulysses in-situ measurements, respectively. Since the
spatial density of grains represented by one of the simulated grain
sizes at large heliocentric distances is not known, the normalization
of the spatial distribution is a free parameter that can be scaled by
an overall factor such that the square of the difference of prediction
and measurement is minimized. If grains with dynamical parameters
similar to the simulated $0.1\;{\rm \mu m}$-, $0.2\;{\rm \mu m}$-, or
$0.3\;{\rm \mu m}$-grains dominate the interstellar dust population
measured by Galileo and Ulysses, the predicted temporal variation of
the flux of the corresponding grain size should match the
measurements.

Galileo has measured an increasing flux of interstellar grains on its
way to Jupiter. According to the simulation this is best explained
with grains that interact strongly with the solar wind magnetic field
and are thus deficient in the inner Solar System. Because Galileo was
moving upwind to larger heliocentric distances, it detected an
increasing flux. The same deficiency of grains in the inner Solar
System can be created by the effect radiation pressure, but to explain
an increase of flux outside $4\;{\rm AU}$ observed by Galileo, the
grains must have very high $\beta$ values of $2.5$ and larger, which
is not supported by the Mie-calculations. Compared to the duration of
the solar cycle, Galileo spent a short time in the interplanetary
space of the outer Solar System. Therefore, the comparison of the
Galileo measurements with the results of the simulation is rather
ambiguous and does not give any information about the effect of the
changing solar cycle on the distribution of interstellar dust grains
in the Solar System.

Since the Ulysses measurements cover nearly $6$ years, the effect of
the changing solar cycle can be observed in the Ulysses data. The
measured flux stayed constant after Ulysses left the ecliptic plane in
February 1992 until mid-1996. After that the flux of interstellar
grains decreased until it reached its minimal value of $(4 \pm 1)\cdot
10^{-5}\;{\rm m}^{-2}\;{\rm s}^{-1}$ at the end of 1997. For
$0.1\;{\rm \mu m}$ grains the simulation does not predict a constant
flux at the Ulysses location between $1992$ and $1996$. Furthermore,
the flux of the simulated $0.1\;{\rm \mu m}$-grains decreases by an
order of magnitude within one year after mid-1996. This is also not
observed. Therefore, the in-situ measurements as well as the
simulation indicates that grains with charge-to-mass ratios of
$5\;{\rm C}\;{\rm kg}^{-1}$ or larger do not contribute significantly
to the flux of interstellar grains in the Solar System within
Jupiter's orbit. For $0.2\;{\rm \mu m}$- and $0.3\;{\rm \mu m}$-grains
(charge-to-mass ratios between $0.5\;{\rm C}\;{\rm kg}^{-1}$ and
$1.4\;{\rm C}\;{\rm kg}^{-1}$) the simulation predicts a flux
variation that is in good agreement with the observation, though
long-term measurements are necessary to confirm the
agreement. Interstellar grains with lower charge-to-mass ratios do not
show a decrease in flux (see right panel of figure
\ref{fig_modelfluxes}). This explains why bigger grains dominate the
mass distribution of interstellar grains measured by Ulysses after
mid-1996 compared to the mass distribution measured between 1992 and
mid-1996 [\markcite{{\it Landgraf et al.,} 1999}].

\section{Conclusion}
The model described in section \ref{sec_modeldescription} explains the
observed time variability of the flux and grain mass distribution of
interstellar dust in the Solar System as due to electromagnetic
interaction of the grains with the changing solar wind magnetic
field. It also shows that very small grains with large charge-to-mass
ratios are repelled from the Sun by the average electromagnetic
effects. Since the grain size distribution in the LIC is believed to
be very steep, the most abundant interstellar grains have
charge-to-mass ratios just small enough (or, equivalently, sizes large
enough) for their inertia to overcome the electromagnetic
repulsion. In this marginal case the electromagnetic perturbation is
still strong and defines the spatial distribution of the grains.  The
knowledge of the spatial distribution of interstellar dust grains can
be used to predict in-situ measurements by spacecraft (e.g. Cassini
and Stardust) [\markcite{{\it Landgraf and M\"uller}, 1998}] depending
on their location in the
Solar System.

The simulation has furthermore shown that the spatial distribution of
small interstellar dust grains deviates strongly from the distribution
of bound, interplanetary grains. This asymmetry as well as the time
variability of the spatial distribution can be used to observe the
infrared emission of interstellar dust in the Solar System by
telescopic means and distinguish it from the radiation emitted by
interplanetary grains. Because of the low average spatial density (about
$10^{-15}\;{\rm cm}^{-3}$ to $10^{-14}\;{\rm cm}^{-3}$) of
interstellar grains, an instrument of very high sensitivity is needed for
such an observation.

Further long-term in-situ measurements are needed to confirm the time
variation of the interstellar dust flux discovered in the Ulysses
data. The simulation has shown that even small interstellar dust
grains reach heliocentric distances of $1\;{\rm AU}$. Therefore, an
Earth orbiting satellite can be used to measure or even collect dust
grains from outside the Solar System.

\acknowledgements 
This work was performed while ML held a National
Research Council-NASA/JSC Research Associateship and is based on ML's
PhD thesis ``Modellierung der Dynamik und Interpretation der
In-Situ-Messung interstellaren Staubs in der lokalen Umgebung des
Sonnensystems'', Ruprecht--Karls--Universit\"at Heidelberg, 1998.

\clearpage
\clearpage
\begin{figure}[ht]
\begin{tabular}{cc}
\epsfxsize=.9\hsize
\epsfxsize=.9\hsize
\epsfxsize=.9\hsize
\epsfxsize=.9\hsize
\end{tabular}
\caption{\label{fig_dd0705} The spatial distribution of $0.7\;{\rm \mu
m}$- (upper two panels) and $0.5\;{\rm \mu m}$- (lower two panels)
grains in a $20\;{\rm AU}\times 20\;{\rm AU}$ area in the
$x$-$y$-plane. The linear color scale represents the spatial grain
density in units of the density at infinity. The two panels on the
left have been calculated using the numerical simulation of particle
trajectories and the two panels on the right show the result of a
analytical calculation for $q/m = 0$ but the same $\beta$-values (see
table~2). The numerical simulation is in good agreement with the
results from the analytical calculation, the deviations result from
the limited number of particles and the limited spatial resolution of
the simulation. As expected, gravitational focusing is weaker for
$0.5\;{\rm \mu m}$-grains due to their higher $\beta$-value.}
\end{figure}
\clearpage
\clearpage
\begin{figure}[ht]
\epsfysize=.8\vsize
\caption{\label{fig_dd02_xy} Distribution of simulated interstellar
$0.2\;{\rm \mu m}$ grains in the $x$-$y$-plane during the solar cycle
between 1991 and 2012. The time sequence starts at the upper left
panel and ends at the lower right. Time steps are about 2.5
years. Plots are $80\;{\rm AU}\times 80\;{\rm AU}$}.
\end{figure}
\clearpage
\begin{figure}[ht]
\epsfysize=.8\vsize
\caption{\label{fig_dd02_xz} Distribution of simulated interstellar
$0.2\;{\rm \mu m}$ grains in the $x$-$z$-plane during the solar cycle
between 1991 and 2012. The time sequence starts at the upper left
panel and ends at the lower right. Time steps are about 2.5
years. Plots are $80\;{\rm AU}\times 80\;{\rm AU}$.}
\end{figure}
\clearpage
\begin{figure}[ht]
\epsfysize=.8\vsize
\caption{\label{fig_dd01_xy} Distribution of simulated interstellar
$0.1\;{\rm \mu m}$ grains in the $x$-$y$-plane during the solar cycle
between 1991 and 2012. The time sequence starts at the upper left
panel and ends at the lower right. Time steps are about 2.5
years. Plots are $80\;{\rm AU}\times 80\;{\rm AU}$.}
\end{figure}
\clearpage
\begin{figure}[ht]
\epsfysize=.8\vsize
\caption{\label{fig_dd01_xz} Distribution of simulated interstellar
$0.1\;{\rm \mu m}$ grains in the $x$-$z$-plane during the solar cycle
between 1991 and 2012. The time sequence starts at the upper left
panel and ends at the lower right. Time steps are about 2.5
years. Plots are $80\;{\rm AU}\times 80\;{\rm AU}$.}
\end{figure}
\clearpage
\clearpage
\clearpage
\clearpage
\begin{table}
\centering
\begin{tabular}{rcccc}
\tableline
$\epsilon$ & solar activity & \multicolumn{2}{c}{avg. polarity} & year
(approx.)\\
&& North & South &\\
\tableline
\tableline
$-\frac{\pi}{2}$ & max & $0$ & $0$ & $1991$\\
\tableline
$0$ & min & $+$ & $-$ & $1997$ \\
\tableline
$\frac{\pi} {2}$ & max & $0$ & $0$ & $2002$\\
\tableline
$\pi$ & min & $-$ & $+$ &
$2008$\\
\tableline
$\frac{3}{2}\pi$ & max & $0$ & $0$ & $2013$\\
\tableline
\label{tab_cycle} 
\end{tabular}
\caption{Average polarity phases of the solar wind magnetic
field in the $22$-year solar cycle parameterized by the current sheet
tilt angle $\epsilon$. If the north/south polarity configuration is
+/-, the magnetic north pole is the heliographic north pole, and for
-/+ the magnetic north pole is the heliographic south pole.}
\end{table}
\vspace{3cm}
\begin{table}[ht]
\centering
\begin{tabular}{*{3}{r@{.}l}cc}
\tableline
\multicolumn{2}{c}{radius $a$ $[{\rm \mu m}]$} &
\multicolumn{2}{c}{$\beta$} &
\multicolumn{2}{c}{$\frac{q}{m}$ $[{\rm C}\;{\rm kg}^{-1}]$} &
mass $m$ $[{\rm kg}]$ & charge $q$ $[{\rm C}]$\\
\tableline
$0$&$7$ & $0$&$46$ & $0$&$108$ & $3.6\cdot 10^{-15}$ & $3.9\cdot
10^{-16}$\\
$0$&$5$ & $0$&$70$ & $0$&$213$ & $1.3\cdot 10^{-15}$ & $2.8\cdot
10^{-16}$\\
$0$&$3$ & $1$&$1$ & $0$&$590$ & $2.8\cdot 10^{-16}$ & $1.7\cdot
10^{-16}$\\
$0$&$2$ & $1$&$4$ & $1$&$33$ & $8.4\cdot 10^{-17}$ & $1.1\cdot
10^{-16}$\\
$0$&$1$ & $1$&$2$ & $5$&$31$ & $1.0\cdot 10^{-17}$ & $5.3\cdot
10^{-17}$\\
\tableline
\end{tabular}
\caption{\label{tab_dynpara}Dynamic parameters of the simulated grain
sizes. The mass is determined by assuming the grains to be homogeneous
spheres with a density of $2.5\;{\rm g}\;{\rm cm}^{-3}$. The radiation
pressure parameter $\beta$ is taken from Mie-calculations for spheres
made of astronomical silicates [\markcite{{\it Gustafson,} 1994};
\markcite{{\it Draine and Lee,} 1984}], and the charge is calculated
for a sphere with a surface potential of $U=+5\;{\rm V}$.}
\end{table}

\clearpage
\end{article}
\end{document}